\documentclass[prl, preprintnumbers ,showpacs,amsmath,amssymb, superscriptaddress,twocolumn]{revtex4-1}
\usepackage{bm}
\usepackage{amsmath}
\usepackage{amssymb}
\usepackage{amsthm}
\usepackage{amsfonts}
\usepackage{enumerate}
\usepackage{latexsym}
\usepackage{times}
\usepackage{ulem}
\usepackage{graphicx}
\usepackage{color}
\usepackage{makeidx}
\usepackage{ifpdf}

\begin{document}

\title{Metal-insulator transition and orbital reconstruction in Mott quantum wells of NdNiO$_{3}$}

\author{Jian~Liu} \email{jxl026@uark.edu}
\affiliation{Department of Physics, University of Arkansas, Fayetteville, Arkansas 72701, USA}
\affiliation{Advanced Light Source, Lawrence Berkeley National Laboratory, Berkeley, California 94720, USA}
\author{M.~Kareev}
\affiliation{Department of Physics, University of Arkansas, Fayetteville, Arkansas 72701, USA}
\author{D.~Meyers}
\affiliation{Department of Physics, University of Arkansas, Fayetteville, Arkansas 72701, USA}
\author{B.~Gray}
\affiliation{Department of Physics, University of Arkansas, Fayetteville, Arkansas 72701, USA}
\author{P.~Ryan}
\affiliation{Advanced Photon Source, Argonne National Laboratory, Argonne, Illinois 60439, USA}
\author{J.~W.~Freeland}
\affiliation{Advanced Photon Source, Argonne National Laboratory, Argonne, Illinois 60439, USA}
\author{J.~Chakhalian}
\affiliation{Department of Physics, University of Arkansas, Fayetteville, Arkansas 72701, USA}


\begin{abstract}
  The metal-insulator transition (MIT) and the underlying electronic and orbital structure in $e_{g}^{1}$ quantum wells based on NdNiO$_{3}$ was investigated by d.c. transport and resonant soft x-ray absorption spectroscopy. By comparing quantum wells of the same dimension but with two different confinement structures, we explicitly demonstrate that the quantum well boundary condition of correlated electrons is  critical to selecting the many-body ground state. In particular, the long-range orderings and the MIT are found to be strongly enhanced under quantum confinement by sandwiching NdNiO$_{3}$ with the wide-gap dielectric LaAlO$_{3}$, while they are suppressed when one of the interfaces is replaced by a surface (interface with vacuum). Resonant spectroscopy reveals that the reduced charge fluctuations in the sandwich structure are supported by the enhanced propensity to charge ordering due to the suppressed $e_g$ orbital splitting when interfaced with the confining LaAlO$_{3}$ layer.
\end{abstract}

\maketitle

Low-dimensional electronic systems have been a focus of condensed matter physics for decades and are known to exhibit numerous fascinating quantum states. A prominent example is quantum well based on semiconductor heterojunctions which has become a foundation of modern technology. Recently, artificial confinement of Mott electrons is emerging as a compelling route to create new two-dimensional (2D) systems unobtainable in the bulk form \cite{Bednorz,Mannhart}. Heterostructures involving \textit{i.e}. cuprates, manganites and LaAlO$_3$/SrTiO$_3$ have revealed the spectacular versatility and fascinating playground of complex oxide interfaces \cite{Ohtomo,Chakhalian1,Logvenov,Bibes}.

In Mott quantum well structures, an ultrathin electronically active slab of correlated oxide is constrained in an epitaxial heterostructure where the vertical hopping of the $d$-electrons is typically suppressed by  dielectric oxide blocks. This geometry was previously shown to provide competing pathways to polar compensation \cite{Hotta,Jian1}, and has recently been harnessed to generate unusual confinement-induced orbital reconstruction \cite{Seo,Benckiser,Freeland} and dimensionality-controlled metal-insulator transition (MIT) \cite{Jian,Boris,Scherwitzl,Yoshimatsu,Yoshimatsu1}. Appealing many-body phenomena have also been predicted, including proposals of half-metallic semi-Dirac points \cite{Pardo} as well as various engaging orderings of charge, spin and orbital \cite{Chaloupka,Jackeli}.
In particular, quantum confinement of $e^{1}_{g}$-system based on the Ni$^{III}$ ($3d^7$) low-spin state in perovskite nickelate RNiO$_3$ (R=rare earth) is of great interest, due to the possibilities in inducing novel quantum states, \textit{e.g.} topological phases \cite{Yang} and unconventional high-$T_{\rm c}$ superconductivity \cite{Chaloupka,Hansmann,Han}. To realize these intriguing phenomena, a controlled manipulation of the orbital degeneracy of $e_g$ electrons via confinement is required. While confinement-induced orbital polarizations were recently observed in both metallic and insulating states in (001)-oriented superlattices \cite{Benckiser,Freeland}, the quantum well structure has also been shown to affect the ground state from a three-dimensional (3D) correlated metal to a (quasi-)2D Mott insulator with propensity to charge and spin orderings (CO/SO) in both superlattices and ultrathin films \cite{Jian,Boris,Scherwitzl,Kaiser,Gray}. On the other hand, the underlying mechanism of confinement behind this Mott transition is still under debates \cite{Lee}. The key question therefore rests on the fundamental role of the confining interface in modulating the collective orderings in Mott quantum wells beyond affording the blocking boundary.

\begin{figure}[t]\vspace{-0pt}
\includegraphics[width=8.5cm]{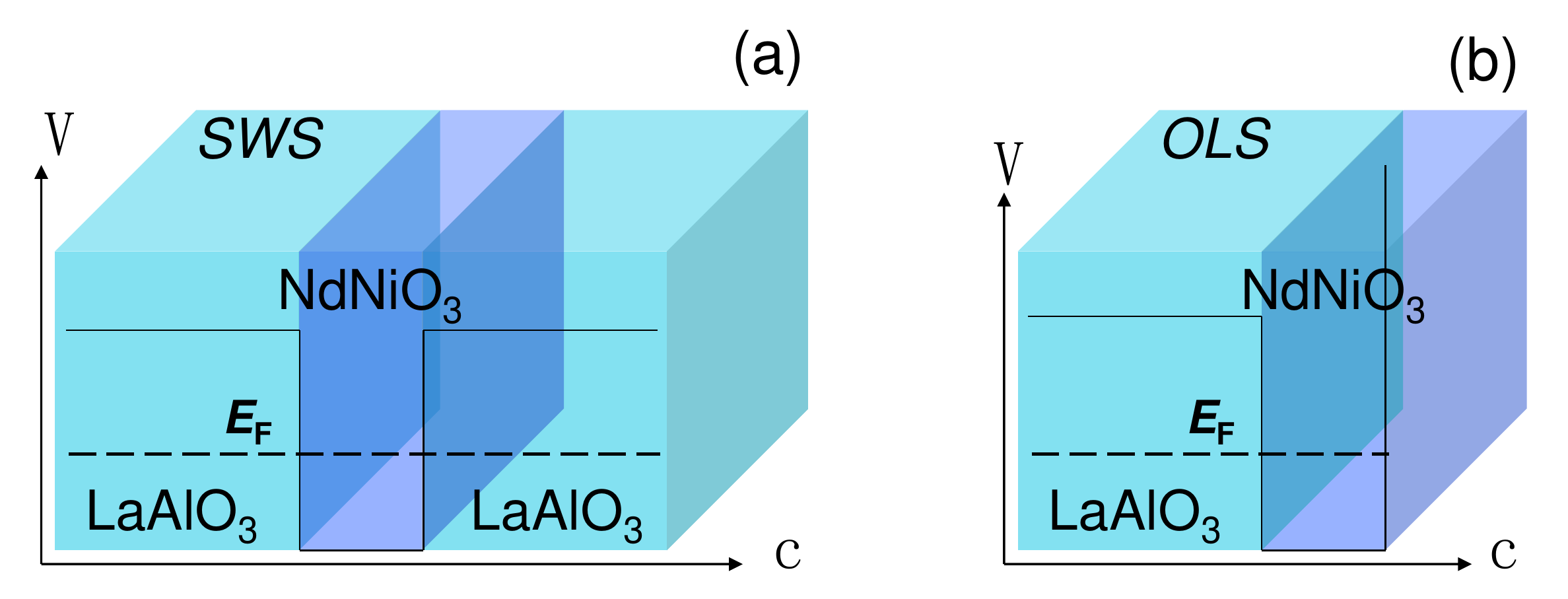}
\caption{\label{wells} (color online) Cartoons of the two quantum well geometries with potential schematics. (a) Confinement by wide-gap dielectrics (LaAlO$_3$) on both sides (sandwiched slab). (b) Confinement by a dielectric (LaAlO$_3$) on one side and vacuum on the other.}
\end{figure}

In this letter, we demonstrate the remarkable capability of the confinement boundary condition on controlling the emerging electronic phase transitions of strongly correlated system in $e^{1}_{g}$ quantum wells consisting of a NdNiO$_3$ active layer.
While previous studies were devoted to LaNiO$_3$, which is a paramagnetic metal in the bulk, NdNiO$_3$ represents an  attractive case of a Mott insulator with cooperative CO and SO that can melt into a correlated metal under increased temperature \cite{Catalan0,Medarde,Staub},  and/or compressive heteroepitaxial strain\cite{Liu,Stewart}.
 Here we report on the dimensional crossover in two types of quantum wells (see Fig.~\ref{wells}) both of which are under identical compressive strain states; the first type involves a sandwiched well structure (SWS) where both sides are interfaced with dielectric LaAlO$_3$, while the other type employs an overlayer structure (OLS) where one of the confining boundaries is replaced by the interface with vacuum (\textit{i.e.} surface).
 Transport measurements demonstrate that, while reducing dimensionality can introduce a MIT in both confinement geometries, the transition temperature and the associated charge- and spin-ordered insulating ground state are remarkably enhanced in SWS compared with OLS. Polarization-dependent local X-ray spectroscopy has identified the presence of an orbital reconstruction between the two geometries, indicating that the fundamentally different response between OLS and SWS is caused by the orbital degeneracy-induced CO instability that couples with the MIT.


To create the quantum wells, high-quality epitaxial NdNiO$_3$ ultrathin slabs oriented along the pseudo-cubic (001) direction were grown by laser molecular beam epitaxy, as described elsewhere in detail \cite{Liu,Meyers}. \textit{In situ} reflection High Energy Electron Diffraction was used to monitor the layer-by-layer growth for `digital' control of the quantum well structure. The compressive strain was achieved via the -0.3\% epitaxial lattice mismatch with LaAlO$_3$ single crystal substrates ($5\times5$ mm$^2$) \cite{Liu}. While ultrathin films of NdNiO$_3$ correspond to the LaAlO$_3$/NdNiO$_3$/vacuum OLS, additional 3-unit-cell (u.c.) slabs of LaAlO$_3$ were added on the top to achieve the LaAlO$_3$/NdNiO$_3$/LaAlO$_3$ SWS. The evolution of the ground state was studied by d.c. transport measurements from 300 {\rm K} to 2 K  in a physical properties measurement system (PPMS, Quantum Design) using the $van$ $der$ $Pauw$ method where the four corners are covered by ohmic contacts (see Supplement). To explore the effects of different boundary conditions on the $e_g$ orbital degeneracy, we have performed detailed resonant soft X-ray absorption measurements at 300 K at the Ni $L$-edge at the 4ID-C beamline of the Advanced Photon Source \cite{Chakhalian1}. The local and element-specific characters of this technique are ideally suited for investigating electronic structure of layers buried in a multi-component environment. To precisely determine the charge state, all spectra were aligned to a NiO standard measured simultaneously with the samples. In addition, each spectrum was  normalized to the beam intensity monitored by a gold mesh set in front of the samples. Since the ${3z^{2}-r^{2}}$ and ${x^{2}-y^{2}}$ orbitals have lobes pointing perpendicular and parallel to the confinement $ab$-plane, respectively, their hole states were probed by rotating the linear polarization of the light to the out-of-plane ($E\parallel c$) and in-plane ($E\parallel a,b$) directions, while fixing the sample position with a 20$^o$ grazing angle of the light (70$^o$ to surface normal) to avoid issues related to geometry corrections.

\begin{figure}[t]\vspace{-0pt}
\includegraphics[width=8.5cm]{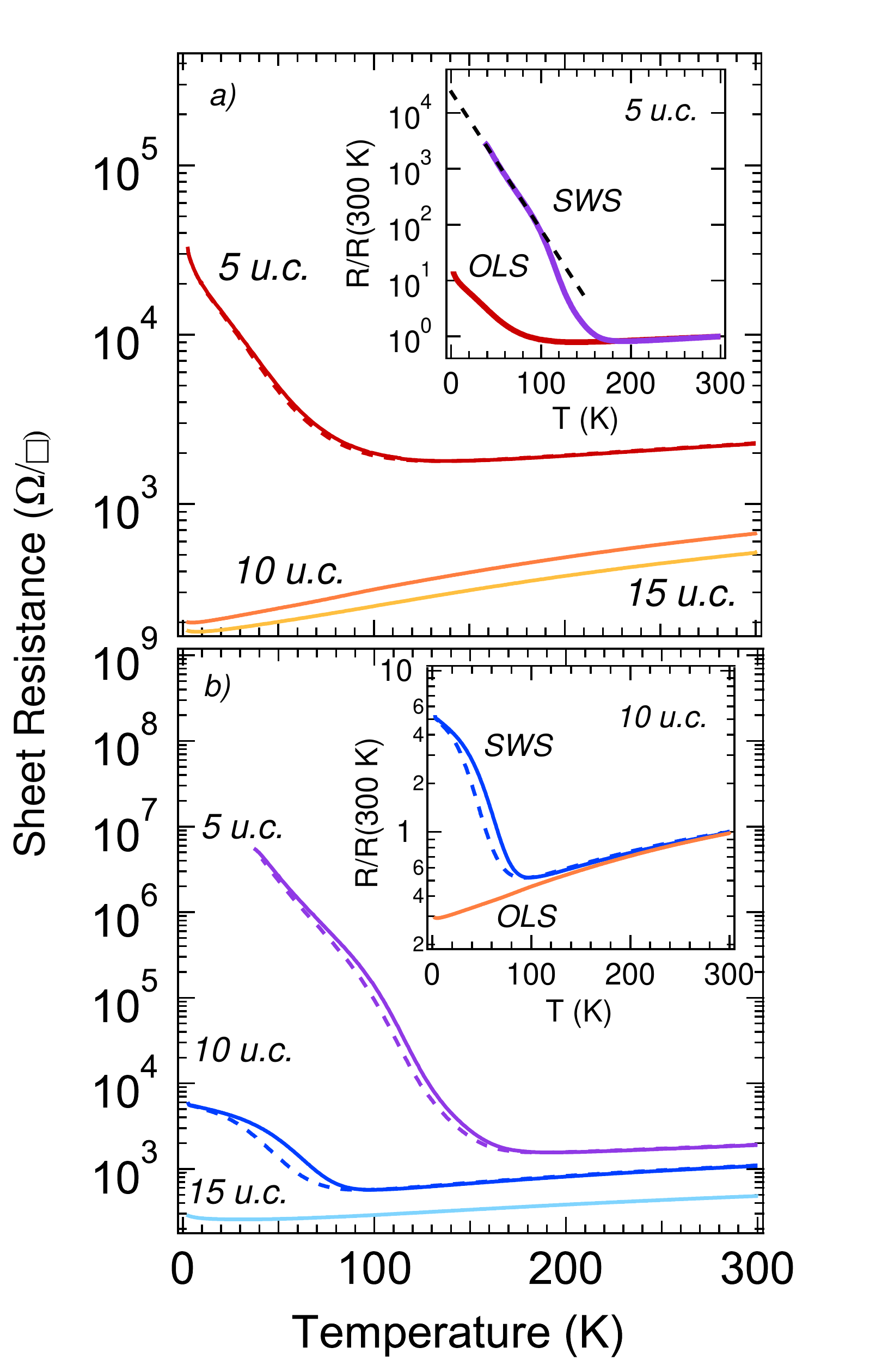}
\caption{\label{transport} (color online) Sheet resistance versus temperature in NdNiO3 quantum wells under OLS (a) and SWS (b). Solid curves are warming data. Cooling data are shown in dashed curves for samples that have hysteresis. Inset of (a): warming curves for 5 u.c. slabs in both geometries. The dashed line is a guide to eye for the extrapolation to lower temperatures in SWS. Inset of (b): comparison of the 10 u.c. OLS and SWS.}
\end{figure}

 Figure~\ref{transport}(a) and (b) show the obtained temperature-dependent sheet resistance of two series of quantum wells of OLS and SWS, respectively, with the NdNiO$_3$ slab thicknesses of 15, 10, and 5 u.c.. As seen, while metallicity is well maintained at high temperatures in all samples, the OLS and SWS display distinctly different low-temperature evolutions upon reducing the confining dimension.
 Specifically, the 15-u.c. OLS remains metallic at low temperatures, which is expected as mentioned earlier due to the complete suppression of the bulk charge- and spin-ordered insulating ground state by the compressive heteroepitaxial strain \cite{Liu,Stewart}. A similar behavior is observed in the 15 u.c. SWS, indicating that quantum confinement is playing a secondary role for thicker slabs, \textit{i.e.} in the 3D regime.
 However, upon shrinking the quantum well to 10 u.c., while no significant change occurs to the low-temperature behavior in OLS, a MIT clearly emerges in SWS at $\sim$100 K below which the resistance raises gradually with a sizable thermal hysteresis, manifesting a strong first order character of the transition. These observations mark the striking difference between the two confinement structures as they are compared in the inset of Fig.~\ref{transport}(b).
 Upon further reduction of  the NdNiO$_3$ slab thickness to 5 u.c., the enhanced quantum confinement facilitates an insulating ground state in OLS, due to the increased correlation/effective mass \cite{Okamoto}. As seen in Fig.~\ref{transport}(a), below $\sim$140 K, the resistance becomes gradually increasing with decreasing temperature accompanied by a small thermal hysteresis between cooling and warming, characteristic of a weak first order MIT. On the other hand, Fig.~\ref{transport}(b) shows that the insulating ground state is radically strengthened in SWS; the onset of the first order transition increases towards $\sim$200 K, below which the resistance is drastically raised by orders of magnitude and eventfully  reaching the   measurement limit below 40 K, denoting a strong MIT. The sharp contrast with the 5 u.c. slab in OLS can also be readily seen in the inset of Fig.~\ref{transport}(a) when the resistance data taken on  warming  are plotted together. An extrapolation for the 5 u.c. SWS suggests that the resistance jump differs by more than three orders of magnitude from that of the OLS.

From the observations described above, one can clearly see that, while both types of confinement structures undergo a dimensionality-controlled MIT, the correlated electrons in NdNiO$_3$ quantum wells of OLS and SWS display drastically different behavior in proximity to the 2D regime. These results unambiguously demonstrate that, when coupled with quantum confinement, the  interface has tremendous effect on driving the MIT and the underlying collective long-range ordering.
In connection to  this, it is interesting to note that due to the reduced coordination of a top atomic layer, the surface is usually believed to have enhanced correlations \cite{Liebsch}. In variance, this surface effect is ruled out by the suppressed MIT in OLS. Since NdNiO$_3$ represents an iconic example of synergetic CO and SO \cite{Catalan0,Medarde}, the suppression/enhancement of these collective electronic behaviors in the quantum well regime  accentuates the pivotal role of the confining boundary.
 Specifically, the strong first order MIT in SWS implies that the ordered insulating phase is favored by the interface with LaAlO$_3$. The surface, on the other hand, effectively suppresses these orderings and the first order character of the transition in OLS, and opposes the carrier localization driven by confinement-induced correlations as the electrons remain relatively mobile. Given that the NdNiO$_3$ slab interfaces with LaAlO$_3$ on the other side, the effect of the surface in destabilizing CO and SO is conceivably much stronger than that in OLS. Moreover, note that the appearance of the first order MIT at the dimensional crossover in NdNiO$_3$ quantum wells is in sharp contrast with the observed second order MIT and N\'{e}el transition in confined LaNiO$_3$ \cite{Jian,Scherwitzl,Boris}. The cause of such difference between these two $e^{1}_{g}$-systems will be an interesting subject for future study.

\begin{figure}[t]\vspace{-0pt}
\includegraphics[width=8.5cm]{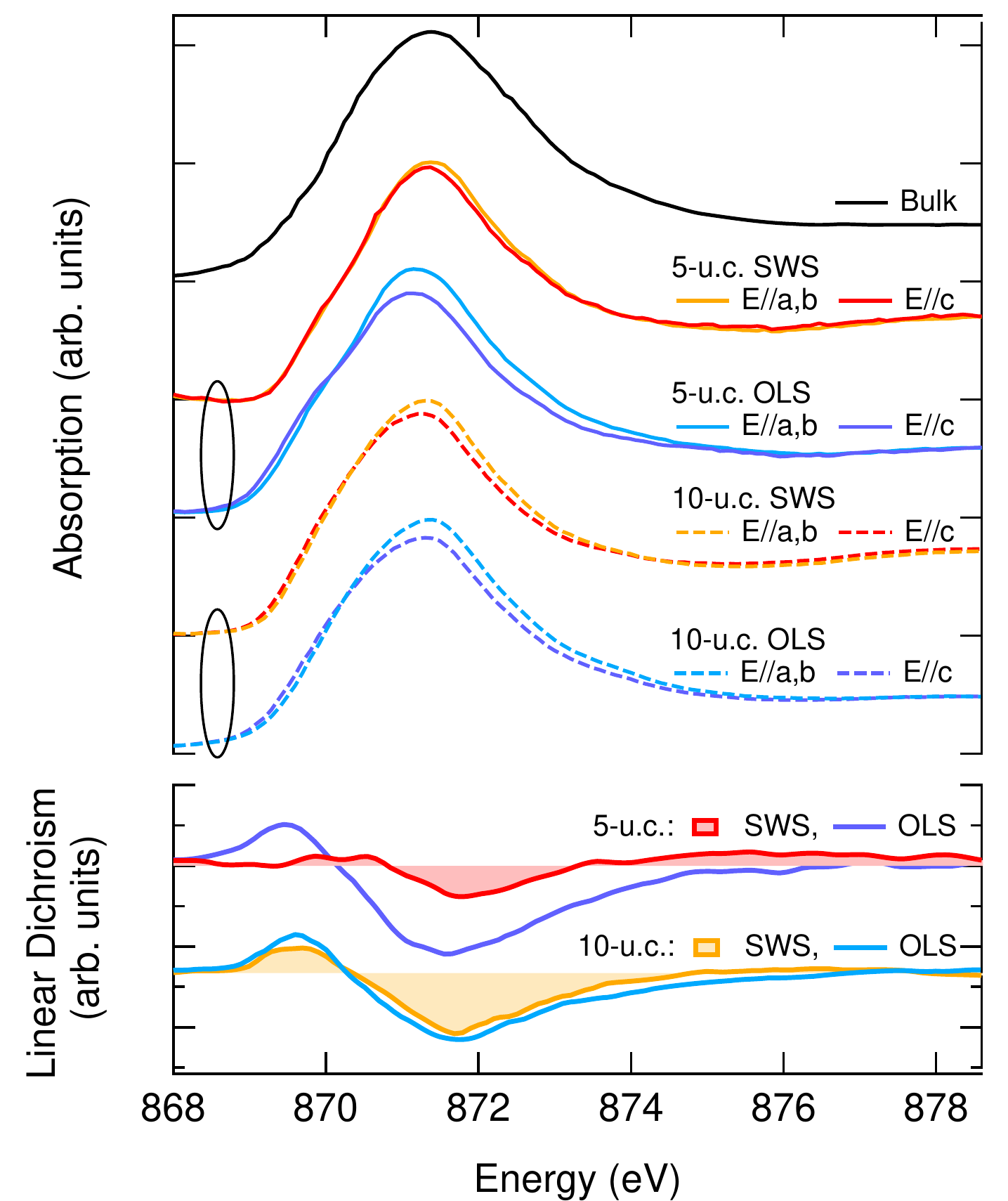}
\caption{\label{NNOL3}
(color online) Polarization-dependent Ni $L$-edge x-ray absorption spectrum (upper) and linear dichroism spectra (lower) ($I_{E\parallel c}-I_{E\parallel a,b}$) of 5-u.c. and 10-u.c. slabs of NdNiO$_3$ in OLS and SWS. Absorption spectrum of bulk NdNiO$_3$ is included for comparison. Note, different samples are compared by normalization to the whiteline intensity at the $L_2$-edge.}
\end{figure}

While the spin degeneracy of localized $d$-electrons is lifted via SO, CO in single valent systems like RNiO$_3$ is to remove the $e_g$ orbital degeneracy via charge disproportionation (e.g. 2Ni$^{III}$ $\longrightarrow$ Ni$^{II}$ + Ni$^{IV}$), instead of the typical Jahn-Teller route \cite{Mazin}. In addition, the orbital physics has also been suggested to be crucial in realizing the cuprate-like behavior in heterostructured $e_{g}^{1}$-systems \cite{Chaloupka,Han}. To gain further microscopic insight into the orbital response to different boundary conditions and avoid complication from the low-temperature orderings, we have performed linearly polarized X-ray absorption measurements above the MITs.
We focused on the 5-u.c. quantum wells, where the effects of the quantum confinement and the boundary condition are strongly coupled, along with bulk NdNiO$_3$ included for direct comparison. First, we note, that for the given small thicknesses of the NdNiO$_3$ slab ($\sim 2$ nm) and the top LaAlO$_3$ slab ($\sim$1.2 nm), the electron-yield probing depth ($\gtrsim$12 nm) is significantly larger than the total film thicknesses of both OLS and SWS; thus, in either case, while the NdNiO$_3$ slab can be assumed to be uniformly probed, the contribution from the top interfacial/surface region will be significant, due to the small total thicknesses. Moreover, the Ni $L_3$-edge is overlapped by the intense La $M_4$-edge from LaAlO$_3$ (see Supplement), rendering the lineshape distortion.
From the distortion-free $L_2$-edge (see Fig.~\ref{NNOL3}),  however, one can observe that the spectra of the NdNiO$_3$ quantum wells taken at 300 K are similar to those of the bulk \cite{Medarde1, Piamonteze}, indicative of a Ni$^{3+}$ charge state of the low-spin configuration (S=1/2). In particular, the spectra consist of a main peak at $\sim$871.2 eV and a weak shoulder at the low-energy side due to the multiplet structure.
This spectral lineshape can be well described by the strong mixing of the $d^{7}$ and $d^{8}\underline{L}$ configurations (here $\underline{L}$ denotes a ligand hole), due to the self-doping caused by the small/negative charge transfer energy \cite{Mizokawa,Mizokawa1}. Meantime, a direct comparison of the spectra shows that, while the low-energy shoulder at the $L_2$-edge is smeared in the bulk, this feature emerges stronger in the quantum well regime due to the reduced effective dimension \cite{Jian}.

Next, both OLS and SWS exhibit linear polarization-dependence across both the $L_3$-edge (see Supplement) and the $L_2$-edge (see Fig.~\ref{NNOL3}). As seen at the bottom of Fig.~\ref{NNOL3}, this dependence is further substantiated by the difference spectra ($I_{E\parallel c}-I_{E\parallel a,b}$).  In particular, the difference spectra in OLS and SWS share a similar lineshape. We recap that the presence of X-ray linear dichroism is the signature of charge anisotropy due to reduced local symmetry. The larger negative dichroic spectral weight at the high-energy side of both edges indicates that the overall absorption cross section is smaller with $E\parallel c$, implying a preferential electron (hole) occupation of the ${3z^{2}-r^{2}}$ (${x^{2}-y^{2}}$) orbital in both OLS and SWS. This assignment of orbital polarization is consistent with that of the NiO$_6$ octahedron distorted by compressive strain \cite{Chakhalian2}. On the other hand, one can see that the orbital polarization is substantially larger in OLS than in SWS; the linear dichroic effect in OLS is about three times of that in SWS. The relative suppression of the ${3z^{2}-r^{2}}$ orbital polarization in SWS strongly implies that the confining interface  stabilizes the preferential occupation of a ${x^{2}-y^{2}}$ orbital for the $e_{g}$ electron, opposing the crystal field effect of compressive strain. This effect of the confining block originates from the interfacial interaction between Ni and Al across the apical oxygen \cite{Hansmann,Han}, and has been previously reported for LaNiO$_3$/LaAlO$_3$ superlattices as an alternative for manipulating orbital polarization \cite{Benckiser,Freeland}. In contrast, the absence of the confining layer on one side of the quantum well selects the ${3z^{2}-r^{2}}$ orbital character, acting synergetically with the compressive strain; this observation is in good agreement with the recent density functional theory calculations \cite{Han1}. To further confirm this picture, we performed the same measurement on the 10-u.c. quantum wells, where the boundary effect is expectedly reduced. Indeed, as shown in Fig.~\ref{NNOL3}, the spectra of both structures clearly indicate their ${3z^{2}-r^{2}}$ orbital characters with a much smaller but still observable difference between them. A close examination reveals that the dichroic effect of the SWS is slightly weaker ($\sim25\%$) than that of the OLS, which can be better seen in their difference spectra. These results corroborate that the confining interface (surface) favors the ${x^{2}-y^{2}}$ (${3z^{2}-r^{2}}$) orbital.

Identifying this counteracting effect between SWS and OLS in lifting orbital degeneracy elucidates  the underlying role of the confining boundary in controlling the collectively ordered correlated ground state and the first order MIT of the quantum wells. Specifically, while Jahn-Teller distortion is characteristic of orbitally degenerate $d$-electrons in the strong interaction limit, CO (charge disproportionation) instability coupled with carrier localization has been shown to be a potent competing pathway to lift the orbital degeneracy on the verge of the localized-to-itinerant crossover in charge-transfer compounds \cite{Mazin}. This is due to the energy gain in the Hund's coupling which overcomes the reduced loss in Coulomb interaction. This mechanism is particularly efficient in charge-transfer systems such as RNiO$_3$ where charges fluctuate strongly via $d^{7}\longleftrightarrow d^{8}\underline{L}$ and the charge-disproportionated Ni$^{IV}$ site can be readily stabilized by forming a singlet state from the $d^{7}\underline{L}$ and $d^{8}\underline{L}^{2}$ configurations rather than creating the energetically unfavorable  $d^{6}$ state. In fact, this CO picture has been successfully applied to explain the complex SO in RNiO$_3$ with the absence of orbital ordering \cite{Mizokawa}. The CO instability, however, would be naturally  suppressed when orbital splitting is artificially introduced and/or enhanced \textit{e.g.} in quantum well structures. As a result, in the critical region of the quantum confinement, when the interface compensates the orbital splitting induced by strain, the CO instability initiates the strong MIT in SWS, In contrast, the ordered insulating ground state is rapidly weakened in OLS and the in-plane charge fluctuation $d^{7}_{3z^{2}-r^{2}}\longleftrightarrow d^{8}\underline{L}_{x^{2}-y^{2}}$ becomes more stable, as the orbital polarization is increased. This difference in the propensity to CO instability between OLS and SWS is further amplified by the absence of  Madelung potential from the confining LaAlO$_3$ layer on one side of OLS; as the prerequisite to the $e_g$ orbital degeneracy, the $d^{7}$ configuration is stabilized by the Madelung potential.

In conclusion, we have investigated the effect of different confinement geometries on the MIT and the orbital reconstruction of correlated electrons in NdNiO$_3$ quantum wells. Direct comparison between OLS and SWS under the stringent dimensionality-control explicitly demonstrates that the interface plays a critical role in defining the many-body ground state of quantum-mechanically confined strongly interacting system, in addition to blocking the vertical hopping. While charge fluctuations are more favorable under confinement with the surface, the ordered insulating ground state is found to be markedly strengthened by interfacing with LaAlO$_3$, resulting in strong first order MITs. Orbital reconstruction was  found between the two confinement structures by resonant X-ray spectroscopy combined with linear polarization-dependence study. The difference between OLS and SWS in triggering the first order MIT was revealed to be caused by the increased (suppressed) the $e_g$ orbital polarization which reduces (enhances) the CO instability in the critical region of the Mott transition induced by quantum confinement. These findings illustrate how the fine balance between competing many-body states of Mott electrons in correlated quantum wells can be  tuned by modulating the  boundary condition.
It is conceivable that such  structures will be useful to exploit these properties in innovative electronic devices.

The authors acknowledge fruitful discussions with M. J. Han, M. van Veenendaal, J. M. Rondinelli, S. Okamoto, D. I. Khomskii, A. J. Millis and G. A. Sawatzky. J.C. was supported by DOD-ARO under the grant No. 0402-17291 and NSF grant No. DMR-0747808. Work at the Advanced Photon Source, Argonne is supported by the U.S. Department of Energy, Office of Science under grant No. DEAC02-06CH11357.


\begin{thebibliography}{99}
\bibitem{Bednorz}
J. G. Bednorz, Nature Mater. \textbf{6}, 821 (2007).

\bibitem{Mannhart}
J. Mannhart and D. G. Schlom, Science \textbf{327}, 1607 (2010).

\bibitem{Ohtomo}
A. Ohtomo and H.Y. Hwang, Nature \textbf{427}, 423 (2004).

\bibitem{Chakhalian1}
J. Chakhalian
et al.
, Science \textbf{318}, 1114 (2007).

\bibitem{Logvenov}
G. Logvenov, A. Gozar and I. Bozovic, Science \textbf{326}, 699 (2009).

\bibitem{Bibes}
M. Bibes, J. E. Villegas and A. Barthelemy, Adv. Phys. \textbf{60}, 5 (2011).

\bibitem{Hotta}
M. Takizawa
et al.
, Phys. Rev. Lett. \textbf{102}, 236401 (2009).

\bibitem{Jian1}
Jian Liu et al., Appl. Phys. Lett. \textbf{96}, 133111 (2010).

\bibitem{Seo}
S. S. A. Seo
et al.
, Phys. Rev. Lett. \textbf{104}, 036401 (2010).

\bibitem{Benckiser}
E. Benckiser et al., Nature Mater. \textbf{10}, 189 (2011).

\bibitem{Freeland}
J. W. Freeland
et al.
, EPL \textbf{96}, 57004 (2011).


\bibitem{Yoshimatsu}
K. Yoshimatsu et al., Phys. Rev. Lett. \textbf{104}, 147601 (2010).

\bibitem{Yoshimatsu1}
K. Yoshimatsu et al., Science \textbf{333}, 319 (2011).

\bibitem{Jian}
Jian Liu et al., Phys. Rev. B \textbf{83}, 161102(R) (2011).

\bibitem{Boris}
A. V. Boris et al., Science \textbf{332}, 937 (2011).

\bibitem{Scherwitzl}
R. Scherwitzl
et al., Phys. Rev. Lett. \textbf{106}, 246403 (2011).

\bibitem{Pardo}
V. Pardo and W. E. Pickett, Phys. Rev. Lett. \textbf{102}, 166803 (2009).

\bibitem{Chaloupka}
J. Chaloupka and G. Khaliullin, Phys. Rev. Lett. \textbf{100}, 016404 (2008).

\bibitem{Jackeli}
G. Jackeli and G. Khaliullin, Phys. Rev. Lett. \textbf{101}, 216804 (2008).

\bibitem{Yang}
A. Ruegg and G. A. Fiete, Phys. Rev. B \textbf{84}, 201103(R) (2011); K.-Y. Yang et al., \textit{ibid.} \textbf{84}, 201104(R) (2011).

\bibitem{Hansmann}
P. Hansmann et al., Phys. Rev. Lett. \textbf{103}, 016401 (2009).

\bibitem{Han}
M. J. Han, C. A. Marianetti, and A. J. Millis, Phys. Rev. B \textbf{82}, 134408 (2010).

\bibitem{Kaiser}
A. M. Kaiser et al., Phys. Rev. Lett. \textbf{107}, 116402 (2011).

\bibitem{Gray}
A. X. Gray et al., Phys. Rev. B \textbf{84}, 075104 (2011).

\bibitem{Lee}
P. Hansmann et al., Phys. Rev. B \textbf{82}, 235123 (2010); S. B. Lee, R. Chen, and L. Balents, \textit{ibid.} \textbf{84}, 165119 (2011).


\bibitem{Catalan0}
G. Catalan, Phase Transitions \textbf{81}, 729 (2008).

\bibitem{Medarde}
M. L. Medarde, J. Phys.: Condens. Matter \textbf{9}, 1679 (1997).

\bibitem{Staub}
U. Staub et al., Phys. Rev. Lett. \textbf{88}, 126402 (2002).

\bibitem{Liu}
Jian Liu
et al.
, Appl. Phys. Lett. \textbf{96}, 233110 (2010).

\bibitem{Meyers}
D. J. Meyers et al., arXiv:1112.5348.

\bibitem{Stewart}
M. K. Stewart et al., Phys. Rev. Lett. \textbf{107}, 176401 (2011).


\bibitem{Okamoto}
S. Okamoto, Phys. Rev. B \textbf{84}, 201305(R) (2011).

\bibitem{Liebsch}
A. Liebsch, Phys. Rev. Lett. \textbf{90}, 096401 (2003).

\bibitem{Mazin}
I. I. Mazin et al., Phys. Rev. Lett. \textbf{98}, 176406 (2007).

\bibitem{Medarde1}
M. Medarde
et al.
, Phys. Rev. B \textbf{46}, 14975 (1992).

\bibitem{Piamonteze}
C. Piamonteze
et al.
, Phys. Rev. B \textbf{71}, 020406(R) (2005).

\bibitem{Mizokawa}
T. Mizokawa, D. I. Khomskii, and G. A. Sawatzky, Phys. Rev. B \textbf{61}, 11263 (2000).

\bibitem{Mizokawa1}
T. Mizokawa et al.
, Phys. Rev. B \textbf{52}, 13865 (1995).

\bibitem{Chakhalian2}
J. Chakhalian
et al.
, Phys. Rev. Lett. \textbf{107}, 116805 (2011).

\bibitem{Han1}
M. J. Han and M. van Veenendaal, Phys. Rev. B \textbf{84}, 125137 (2011).






\end{thebibliography}
\end{document}